# PTNet: A High-Resolution Infant MRI Synthesizer Based on Transformer

Xuzhe Zhang†, *Student Member, IEEE*, Xinzi He†, Jia Guo, Nabil Ettehadi, Natalie Aw, David Semanek, Jonathan Posner, Andrew Laine, *Fellow, IEEE*, and Yun Wang*

*Abstract*—Magnetic resonance imaging (MRI) noninvasively provides critical information about how human brain structures develop across stages of life. Developmental scientists are particularly interested in the first few years of neurodevelopment. Despite the success of MRI collection and analysis for adults, it is a challenge for researchers to collect high-quality multimodal MRIs from developing infants mainly because of their irregular sleep pattern, limited attention, inability to follow instructions to stay still, and a lack of analysis approaches. These challenges often lead to a significant reduction of usable data. To address this issue, researchers have explored various solutions to replace corrupted scans through synthesizing realistic MRIs. Among them, the convolution neural network (CNN) based generative adversarial network has demonstrated promising results and achieves state-of-the-art performance. However, adversarial training is unstable and may need careful tuning of regularization terms to stabilize the training. In this study, we introduced a novel MRI synthesis framework - Pyramid Transformer Net (PTNet). PTNet consists of transformer layers, skip-connections, and multi-scale pyramid representation. Compared with the most widely used CNN-based conditional GAN models (namely pix2pix and pix2pixHD), our model PTNet shows superior performance in terms of synthesis accuracy and model size. Notably, PTNet does not require any type of adversarial training and can be easily trained using the simple mean squared error loss.

*Index Terms* — Infant, Brain, Transformer, MRI synthesis, Performer.

## I. INTRODUCTION

The first two years of life after birth mark the most rapid periods of postnatal growth and development for the human brain. The brain structures, functions, and neural pathways that develop during this time lay the foundation for the people we will become. Therefore, an important goal for many studies of early childhood is identifying early biomarkers of later cognitive functions, behaviors, or risks. Structural magnetic resonance imaging (MRI) has become an important non-invasive approach to investigate brain structural changes at a high spatial resolution. Over the last decade, researchers found a modest relationship between brain structure, cognition, and behavior[1-4], suggesting that with improved methodologies, early imaging biomarkers may be useful in predicting later risk.

Compared with adults, infant brains have 1) lower contrast-to-noise ratios due to the relative lack of myelination and shorter scan times[5]; 2) lower spatial resolution due to the smaller overall volume of the brain; and most importantly 3) tissue intensities change dramatically over the first two years of life. In addition, given infants' characteristics (long preparation time [feeding and swaddling to induce sleep], irregular sleep patterns, and inability to follow instructions to keep still), it is often difficult to collect high-quality multimodal MRI scans for infants before 2 years old[6]. Depending on the research goal of studies and the choice of MRI processing pipelines, researchers tend to often prioritize only one modality. For example, studies with a focus on newborns will most likely prioritize the acquisitions of T2 weighted (T2w) scans in a limited scanning time if they chosen to use developing human connectome project (dHCP) structural pipeline[7] or T1 weighted (T1w) scans if using Infant FreeSurfer pipeline[8]. The above reasons together often lead to low-quality unusable structural MRI scans for single or both modalities. Of note, structural MRI processing (tissue/region segmentation, surface reconstruction) is the first and critical procedure for other modalities' MRI analysis, e.g., functional MRI and diffusion MRI. The consequence of poor quality structural MRI scans can lead to the loss of a subject's entire brain MRI data. Therefore, novel and robust methodologies, which can synthesize missing or corrupted infant MRI scans, can be very useful for developmental neuroscience and clinical research[9].

Previous studies have demonstrated that synthesized single or multimodal MRIs based on existing high quality scans, to some extent, improve biomedical imaging processing procedures, e.g., segmentation and registration. For example, replacing corrupted fluid-attenuated inversion recovery MRI scans with its synthesized version based on corresponding T1w, T2w, and proton density (PD) scans can yield better segmentation[10]. Similarly, a previous study indicated that synthesized T1w scans could replace real T1w scans in inter-modality and cross-subject brain MRI registration, and it outperforms registering with only real PD scans[11].

Prior to the popularity of deep learning (DL), registration-

† indicates equal contribution. * indicates corresponding author.
X. Zhang, X. He, N. Ettehadi, A. F. Laine are with the Department of Biomedical Engineering at Columbia University, New York, NY 10027, USA. J. Guo is with Department of Neurology (in Psychiatry) at Columbia University Irving Medical Center, New York, NY 10032, USA

N. Aw, D. Semanek, J. E. Posner, Y. Wang, are with the Division of Child & Adolescent Psychiatry/Department of Psychiatry at Columbia University Irving Medical Center, New York, NY 10032, USA (e-mail: Yun.Wang @nyspi.columbia.edu)

based and intensity-based transformation methods were prevalent in this domain. Registration-based methods rely on group atlas as well as deformable registration to synthesis image with a different contrast: assuming for the source image $s_1$ with contrast $c_1$, and co-registered atlas scans $a_1$ and $a_2$ with contrasts $c_1$ and $c_2$, respectively, a deformation filed is calculated to non-linearly register $a_1$ to $s_1$, and is then applied to $a_2$, synthesizing the corresponding scan $s_2$ with contrast $c_2$[12]. Although registration-based image synthesis provides promising performance in synthesizing computed tomography and positron emission tomography from MRI[13, 14], it may not be applicable to infant MRI synthesis because of 1) lack of accurate and longitudinal infant brain MRI atlas; 2) more profound variations among infant brain at different ages which may introduce more registration error with the atlas. Intensity-based transformation methods often utilize image analogies, sparse reconstruction, non-linear regression, as well as neural network to achieve image synthesis[10, 11, 15-19]. However, a previous study argued that the above-mentioned methods, either dictionary reconstruction, random forest regression, or neural network-based, tends to lose fine details as well as yield suboptimal synthesis quality[9].

Given the success of generative adversarial network (GAN) in image synthesis, translation, and manipulation[20-23], recent studies have attempted to introduce the convolution neural network (CNN)-based GAN framework into MRI synthesis and shown improved performance compared with aforementioned methods[9, 17, 19, 24]. However, GAN training tends to be unstable and regularization terms (e.g., reconstruction loss, perceptual loss, and cycle-consistency) are often needed. Carefully tuned weights between adversarial loss and regularization loss are necessary for training stability. Recently, the transformer layer, which is a self-attention and convolution-free architecture, has been introduced to the computer vision domain and demonstrates outstanding performance in classification and segmentation in terms of accuracy and efficiency[25, 26]. The performer layer is also introduced and applied to vision tasks[27, 28]; it is a similar attention-based architecture to the transformer but with a simplified self-attention and requires less computation than the transformer.

In this study, we focus on synthesizing infant brain structural MRIs (T1w and T2w scans) using both transformer and performer layers. We design a novel framework, inheriting the U-Net-like as well as multi-resolution pyramid structures[22, 29], and utilizing performer encoder (PE), performer decoder (PD), and transformer bottleneck to synthesize high-quality infant MRI. We conducted several experiments based on a large-scale infant MRI dataset – the Developing Human Connectome Project (dHCP) dataset[30], and compared our model's performance with other methods including pix2pix and pix2pixHD [22, 23]. We demonstrate that our proposed model can synthesize realistic T1w scans based on T2w scans and vice versa. Quantitatively, compared with pix2pix and pix2pixHD models [9, 24], our framework achieves higher structural similarity index measure (SSIM) and peak signal-to-noise ratio (pSNR) when it is validated on the unseen test dataset.

Moreover, our framework only requires a simple mean squared error loss for the training.

## II. RELATED WORKS

### A. GAN-based MRI Synthesis

CNN-based GAN is the most prevalent framework in the image translation and synthesis domain. It utilizes adversarial training, which takes advantage of the feedback from the discriminator network to generate images that are similar to the training dataset. During the training, two subnetworks: generator and discriminator, are trained simultaneously. The generator employs a decoder (original GAN) or an encoder-decoder (conditional GAN) architecture. The original GAN was proposed to unconditionally generate images from latent space noise vector [20]. The discriminator is a classifier trained by the real and synthesized image. The discriminator has access to the true label during the training. The generator is trained using the feedback from the discriminator and aims to fool the discriminator and to generates images that cannot be distinguished from the real images. The conditional GAN has been used in many different downstream applications, such as super-resolution, style transfer, sketch-to-image generation, and image inpainting. However, the training of GAN is usually unstable, especially in the unconditional manner. Stability is slightly improved in the conditional GAN as the input is not random noise but meaningful features extracted by the encoder.

Inspired by the previous success of conditional GANs in natural image translation, previous studies [24] and [9] have used a similar framework in [23] and [22] (namely pix2pix and pix2pixHD), respectively. [9] has explicitly shown that GAN-based methods outperform the previous intensity-based transformation and neural network-based methods (i.e., Replica and Multimodal) in MRI synthesis[17, 19].

### B. Transformer in Computer Vision Tasks

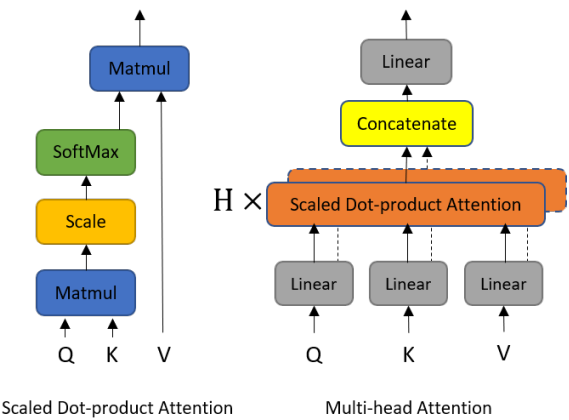

**Fig. 1**. Self-attention mechanism used in Transformer. Head count ($H$) is the number of scaled dot-product attention used in the multi-head attention.

The transformer is an architecture that solely relies on self-attention mechanism (**Fig. 1**) and is completely convolution-free[25]. A transformer layer consists of a multi-head self-attention layer and a fully connected feed-forward network (multilayer perceptron) (**Fig. 2**). A residual connection and a layer normalization are applied in both components. The transformer was originally designed for sequence processing and is becoming a popular and fundamental architecture for NLP tasks. Recently, it has been extended to various computer vision tasks, such as image classification, image segmentation, image generation, object detection[26, 28, 31, 32]. In those applications, the transformer has demonstrated its great potential to achieve or outperform state-of-the-art CNN-based networks, largely because of its self-attention mechanism.

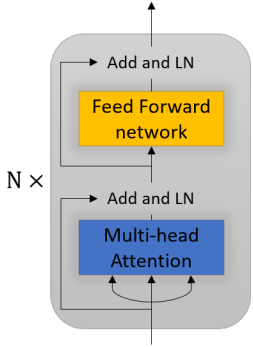

**Fig. 2.** A basic transformer block. Add = residual connection, LN = Layer Normalization.

The self-attention mechanism is based on multiplicative attention through the dot-product of weights and values (of dimension $d_v$, where the weight matrix is calculated by a compatibility function of the query with the corresponding key (of dimension $d_k$). In practice, queries, keys, and values are packed together into a matrix Q, K, V, respectively. The scaled dot-product attention is calculated using (1). Instead of performing the scaled dot-product attention one time, the original paper proposed a multi-head attention (MHA) module[33], which is more beneficial for capturing global dependencies. As shown in **Fig. 1**, Q, K, and V are linearly projected by head-count $H$ times, by linear projections $W^Q$, $W^K$, and $W^V$. For each head, the single head attention is calculated in parallel based on (2). The final output of MHA is given by the linear projection $W^O$ of the concatenation of head attentions as shown in equation (3) below.

$$Attention(Q, K, V) = softmax(\frac{QK^T}{\sqrt{d_k}})V \quad (1)$$

$$head_i(Q, K, V) = Attention(QW_i^Q, KW_i^K, VW_i^V) \quad (2)$$

$$MHA(Q, K, V) = Concat(head_1, \ldots, head_h)W^O \quad (3)$$

### III. METHODS

#### A. Pyramid Transformer Net (PTNet) for MRI Synthesis

In this work, we introduce a novel MRI synthesis framework: Pyramid Transformer Net (PTNet). PTNet's architecture consists of transformer layers, skip-connections, and a multi-scale pyramid representation. An overview of our proposed PTNet model is depicted in **Fig. 3**. Specifically, we introduce the performer-based encoder/decoder to replace the CNN-based encoder/decoder, and adopt the successful structure of U-Net. Though the performer and transformer layers are operated at the level of tokens, we reshape the output features to a matrix at each layer, enabling the skip connection between encoding and decoding paths, aiming to preserve fine structures of the brain. More importantly, we implement a framework with a pyramid structure to: 1) leverage both global and local information during reconstruction, 2) and alleviate the intensive computation need for high-resolution features. The entire framework operates in two levels: the original resolution and downsampled resolution (1/2 in each x and y axis). For each branch, each image is downsampled three times; the result from the low-resolution branch is concatenated back to the high-resolution branch; and the deepest feature map goes through a transformer-based bottleneck.

Our pyramid structure is inspired by several previous studies [34-37]. A pyramid multi-resolution design typically has two branches (**Fig. 3**) and improves vision tasks by aggregating both global and local information. However, the low- and high-resolution branches are usually trained separately and then tuned together, as described in a previous study[22]. In comparison, our framework can be trained end-to-end and simultaneously. In the following sections, we introduce each component of our proposed PTNet model in detail. The performer-based encoder and decoder are introduced in Section A.1, transformer-based bottleneck in Section A.2, and other details in Section B.

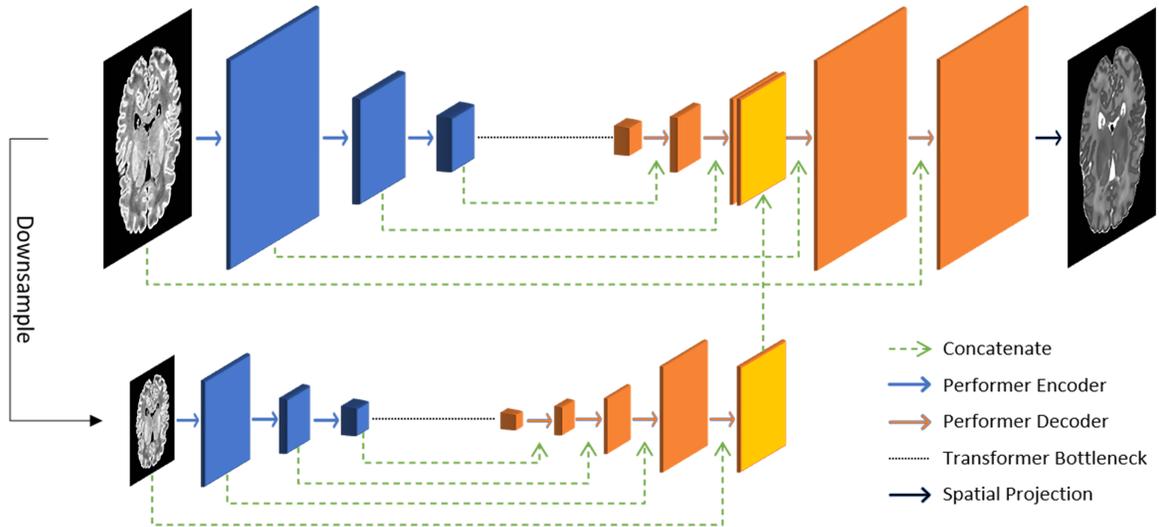

**Fig. 3.** Overview of proposed Pyramid Transformer Net (PTNet). We mimic the classical U-Net structure and inherit the skip connection. We parallelize the conversion at two distinct resolutions and a concatenation between them is applied. The detailed structures of each component are illustrated in the following Fig. 4 and 5. The spatial projection is a fully-connected layer that reduce the channel to output channel number.

*1) Performer Encoder and Decoder*

As introduced in above Section A, the original transformer utilizes the MHA module [33] to better capture global attention. However, such an MHA module is compute-intensive, and its space and time complexities are quadratically grown with input token number (which is proportional to image size in vision tasks). The performer is introduced to estimate and replace the regular SoftMax attention kernel which is applied in the transformer, with linear space and time complexity, as well as provable accuracy. It is achieved by approximating the regular attention kernel by a novel Fast Attention via a positive Orthogonal Random features approach (FAVOR+). Interested readers can find a detailed description in the paper[27].

The most significant challenges for applying the original transformer model are computational time and GPU memories when the input has high spatial resolution, such as in the case of brain MRIs. To solve this issue, instead of the transformer, we adapt the performer in our encoding and decoding blocks and name them as the Performer Encoder (PE) and Performer Decoder (PD), which are illustrated in **Fig. 4**. In PE, for an input tensor with a size of $N \times C_{in} \times X \times Y$, we unfold the 2-D matrix into a series of tokens using a window of $n$ by $n$ and a stride $S$. The resultant tokens are with size $N \times \frac{1}{S^2}XY \times C_{in}n^2$ and are fed into the performer. The output from the performer is then transposed and reshaped to a size of $N \times C_{out} \times X \times Y$ (**Fig. 4 panel a**). During the encoding, the S is often set as 2. In its counterpart PD, the unfolding, performer, transposition, and reshaping remain the same, only an upsampling by a factor of $S$ is applied before the abovementioned steps (**Fig. 4 panel b**).

*2) Transformer Bottleneck*

In the bottleneck, we employed the original transformer layers as the input feature maps, which are already of low spatial size and aim to better capture any global dependencies across the bottleneck features. Such a transformer bottleneck is depicted in **Fig. 5**. The unfolding is similar to PE and we use an $S$ of 2 to further decrease the number of tokens that go through the transformer blocks. After unfolding, a fully connected layer is applied to linearly project the token from $N \times \frac{1}{S^2}XY \times C_{in}n^2$ to $N \times \frac{1}{S^2}XY \times C_{embd}$, where $C_{embd}$ represents the embedding dimension through the transformer blocks. Before feeding the bottleneck embeddings into the transformer blocks, we add the positional encoding, which aims to provide some information about the relative or absolute position of the tokens in the sequence[25]. In this work, we utilize the same *sine* and *cosine* functions to generate the positional encoding according to (4) and (5), where $pos$ is the position and $i$ is the dimension. After that, the embeddings go through $M$ transformer blocks and the output is transposed and reshaped as usual.

$$PE_{pos,2i} = sin(pos/10000^{2i/C_{embd}}) \quad (4)$$

$$PE_{pos,2i+1} = cos(pos/10000^{2i/C_{embd}}) \quad (5)$$

## B. Model Details

We proposed two types of PTNet: PTNet-S (Small) and PTNet-L (Large). They have the same architecture as shown in **Fig. 3** but have different numbers of transformers in their bottlenecks. The PTNet-L employs 9 transformer blocks in both high- and low-resolution branches, while PTNet-S utilizes 1 and 2 transformer blocks, respectively. Except for the first PE and the last PD that have an unfolding window size of 7 ($n = 7$, **Fig. 4**), other PEs, PDs, and transformer blocks have the same $n = 3$. Similarly, only the first PE and the last PD have a stride of 1 during unfolding ($S = 1$, **Fig. 4**), which does not change the spatial resolution of the feature maps, $S$ is set as 2 in all other blocks. The $C_{out}$ is set as 32, 64, 128 for PEs, and is set as 64, 32, 32, 32 for PDs. The header count H in the MHA module is always set as 4 (**Fig. 1**). The embedding dimension $C_{embd}$ of the transformer bottleneck (**Fig. 5**) is set as 256 and 512 for high- and low-resolution branches, respectively. As shown in **Fig. 3**, the input and the output of each PE are saved and concatenated back to the decoding path, and the output of the lower branch is concatenated back into the upper branch. The Spatial Projection linearly projects the output of the last PD from a dimension of 32 to 1, generating the outcome.

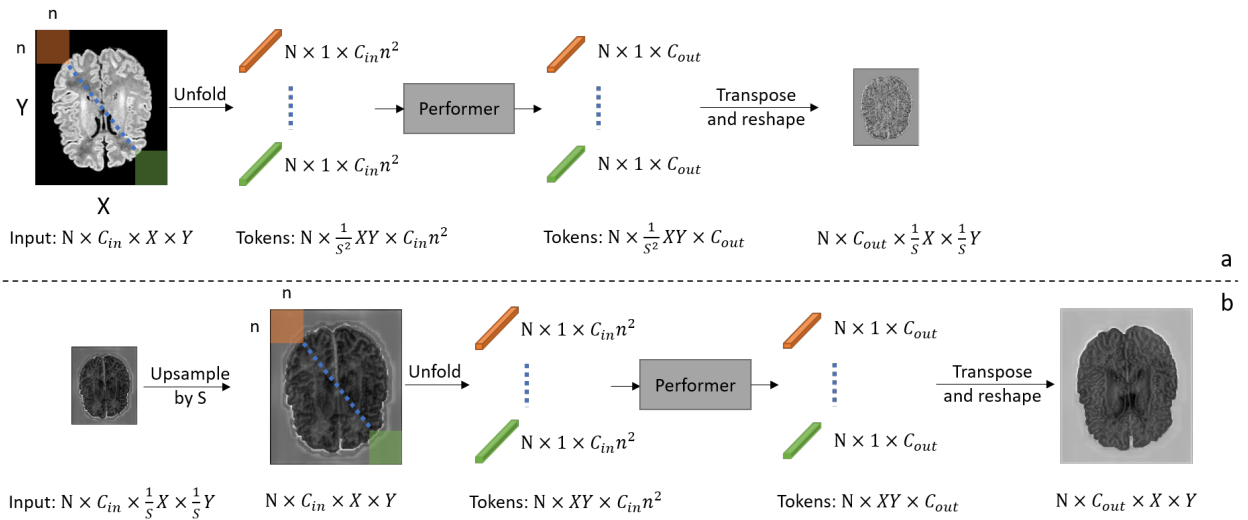

**Fig. 4.** Proposed performer encoder (PE, a) and performer decoder (PD, b). a): The PE will first unfold the feature maps into tokens. The channel after unfolding is decided by the input channel $C_{in}$ and unfold kernel size $n$. Unfolded tokens are then fed into a performer layer. The resultant token is lastly transposed and reshaped to a feature map which has been downsample by a scale of $S$ (stride). b): PD will first upsample the input feature maps by a factor of $S$. The upsampled feature maps are then processed as mentioned in the PE, but there is no stride so the upsampled feature size remains unchanged. The factor $S$ is usually set as 2. $C_{in}$ and $C_{out}$ are changed at different levels of the network. The kernel size $n$ is usually set as 3, and 7 is used when it is in the first layer.

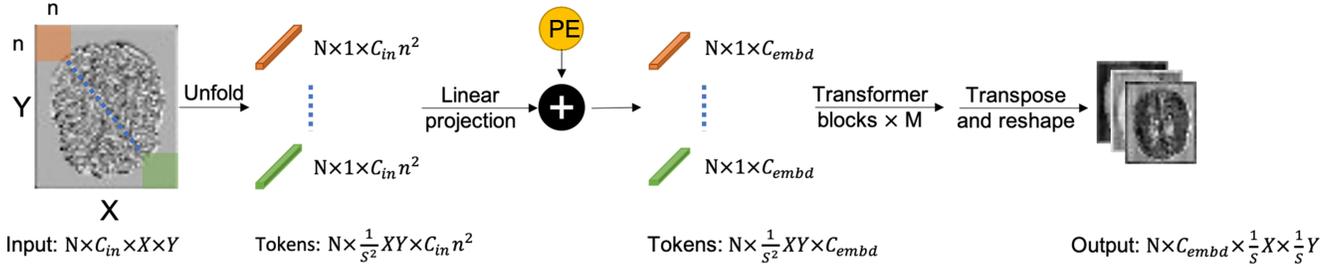

**Fig. 5.** Proposed transformer bottleneck. The unfolding is similar to PE and PD. And additional position embedding and linear projection are used prior to feeding in transformer blocks. The output of M transformer blocks are then transposed and reshaped and feed into PDs.

### C. Datasets

#### 1) dHCP dataset

We used 459 paired T1w and T2w scans from dHCP v1.0.2 data release. The infant structural T1w and T2w scans from dHCP were collected before one month. The quality of skull-stripping and co-registration was assessed by a senior MRI technician. Among them, 43 pairs of scans were excluded because of corrupted T1w scans, poor skull-stripping, or co-registration quality. In total, we include 416 paired T1w and T2w scans and split them into training, validation, and testing sets with a ratio of 7:1:2. The image has an isotropic $0.5 \times 0.5 \times 0.5$ $mm^3$ resolution. The matrix size is $290 \times 290 \times 203$ and is cropped to $256 \times 224 \times 203$ by excluding the all-zero background. To remove outliers with extremely high intensities, each volume is normalized to [0,1] by its minimum intensity value and 99.95 percentile maximum intensity value. Then we resliced each volume into continuous slices, resulting in 59,073 axial slices (from 291 volumes) in the training dataset, 8,526 axial slices (from 42 volumes) in the validation dataset, and 16,849 axial slices (from 83 volumes) in the testing dataset. More details about the image acquisition can be found in the original work[30].

### D. Experiments

We designed several experiments with different components and models as described in **Table 1**. Specifically, three different models were tested and compared, including pix2pix, pix2pixHD, and our proposed PTNet. The pix2pix model is a U-Net-like model and its generator is an encoder-decoder that progressively downsampled the feature maps by a factor of 2 while increasing the dimension of the feature maps. It does not use any bottleneck layers but uses a skip connection between the encoder and decoder path.

The pix2pixHD is an advanced version of pix2pix, which utilized a residual block as the bottleneck layer's backbone. The pix2pixHD-global only employs a single generator proposed by [38]; the pix2pixHD-local is additionally equipped with a local enhancer network that works on high-resolution feature maps[22]. The pix2pixHD-global employs 9 residual blocks in its bottleneck, and its -local version utilized 3 and 6 residual blocks in high- and low-resolution branches, respectively. The PTNet-S and -L denote small and large versions that are equipped with different amounts of transformer blocks in different branches' bottlenecks.

As explained and demonstrated in the previous studies[22, 23], for fair comparisons, we used the same training strategies for three models.

**TABLE 1.** ALL MODELS' DETAILS

| Models | Encoder | Decoder | Down-sample Ratio | Bottleneck layers | |
|---|---|---|---|---|---|
| | | | | Backbone | Block number |
| pix2pix | Conv2D | Conv-Trans2D | 1/32 | N/A | N/A |
| pix2pixHD -global | Conv2D | Conv-Trans2D | 1/16 | ResBlock | 9 |
| pix2pixHD -local | Conv2D | Conv-Trans2D | 1/16 | ResBlock | 3+6 |
| **PTNet-S** | PE | PD | 1/16 | Transformer | 1+2 |
| **PTNet-L** | PE | PD | 1/16 | Transformer | 9+9 |

The PyTorch implementations of pix2pix and pix2pixHD are both publicly available[1,2]. The pix2pix series need both adversarial loss ($L_{adv}$, (6)) and other regularization terms to stabilize the training. Therefore, a generator $G$ and discriminator $D$ are used during the training. We termed $X$ as the input source image, and $Y$ as the target image. For the pix2pix, we used the $L_{adv}$ and $L1$ reconstruction loss (mean absolute error, (7), with a weight of 100) as the loss function described in (8); for the pix2pixHD, instead of incorporating an $L1$ reconstruction loss, we incorporated the $L1$ loss in the feature-level ($L_{feat}$, (9), with a weight of 10) with the $L_{adv}$ (10); for our PTNet, it could be efficiently trained using only a simple mean squared error loss (11). It should be noted that, for (9), $D_i$ indicated the output from the *i-th* layer of the discriminator $D$.

$$L_{adv} = \mathbb{E}_{X,Y}[\log D(X,Y) + \log(1 - D(X, G(X)))] \quad (6)$$

$$L_{MAE} = mean(\sum_x \sum_y |G(X)(x,y) - Y(x,y)|) \quad (7)$$

$$L_p = L_{adv} + 100 * L_{MAE} \quad (8)$$

$$L_{feat} = mean(\sum_i (D_i(X,Y) - D_i(X, G(X))) \quad (9)$$

---
[1.] https://phillipi.github.io/pix2pix/
[2.] https://github.com/NVIDIA/pix2pixHD

$$L_{pH} = L_{adv} + 10 * L_{feat} \quad (10)$$

$$L_{MSE} = mean(\sum_x \sum_y (G(X)(x,y) - Y(x,y))^2) \quad (11)$$

Models were separately trained for T1w-to-T2w and T2w-to-T1w conversions. For T1w-to-T2w conversion, $X$ was T1w scan and $Y$ was the corresponding T2w scan, and vice versa. We trained $G$ to minimize the corresponding loss function (8), (10), or (11). For a fair comparison, all models were trained with a batch size of 4, for 5 epochs with a fixed learning rate $2e^{-4}$ and 5 epochs with a linearly decreasing learning rate (to 0). After training, the model with the highest Structural Similarity Index Measure (SSIM) on the validation dataset was selected for comparison on the testing dataset. All experiments were trained on an NVIDIA GeForce RTX 2080 Ti GPU with 12 GB memory. The entire framework was implemented in PyTorch (and will be available on GitHub soon https://github.com/XuzheZ/PTNet).

To quantitatively compare the performance of different models, the SSIM and pSNR were calculated on test dataset [39, 40]. Instead of assessing the image quality from quantification of intensity difference as pNSR does, SSIM is considered to be correlated to the quality perception of the human visual system in the perspective of distortion and degradation of structural information. The detailed mathematical justification can be found in the original study [39, 40]. In our study, we normalized the ground truth and synthesized volumes to the same intensity range [0,1]. SSIM and PSNR were calculated using (12) and (13), to be noted that, in (12) and (13), $X$ and $Y$ represented volume instead of slice; $\mu$ indicated mean intensity; $\sigma$ was standard deviation; $\sigma_{XY}$ was the covariance between $X$ and $Y$; positive constant $C$ was used to prevent division by zero.

$$PSNR(X,Y) = 10 log_{10}\left(\frac{1}{MSE(X,Y)}\right) \quad (12)$$

$$SSIM(X,Y) = \frac{2\mu_X\mu_Y + C_1}{\mu_X^2 + \mu_Y^2 + C_1} \frac{2\sigma_X\sigma_Y + C_2}{\sigma_X^2 + \sigma_Y^2 + C_2} \frac{\sigma_{XY} + C_3}{\sigma_X\sigma_Y + C_3} \quad (13)$$

We used paired T-test to compare SSIM and pSNR between different methods.

## IV. RESULTS

### A. Visual Comparisons of Synthesized MRI

We first provided visualizations of synthesized scans in three anatomical views (sagittal, axial, and coronal) and their absolute error maps from our proposed and other models (**Fig. 6**). Absolute error is calculated between normalized ground truth and converted scans, ranged from [0,1] – lower values (darker color) indicate more closeness to the real scans. From **Fig 6**., we found that our model produces less absolute error than the GAN-based models in general. Especially, our model produced more realistic synthesis in the circled regions where

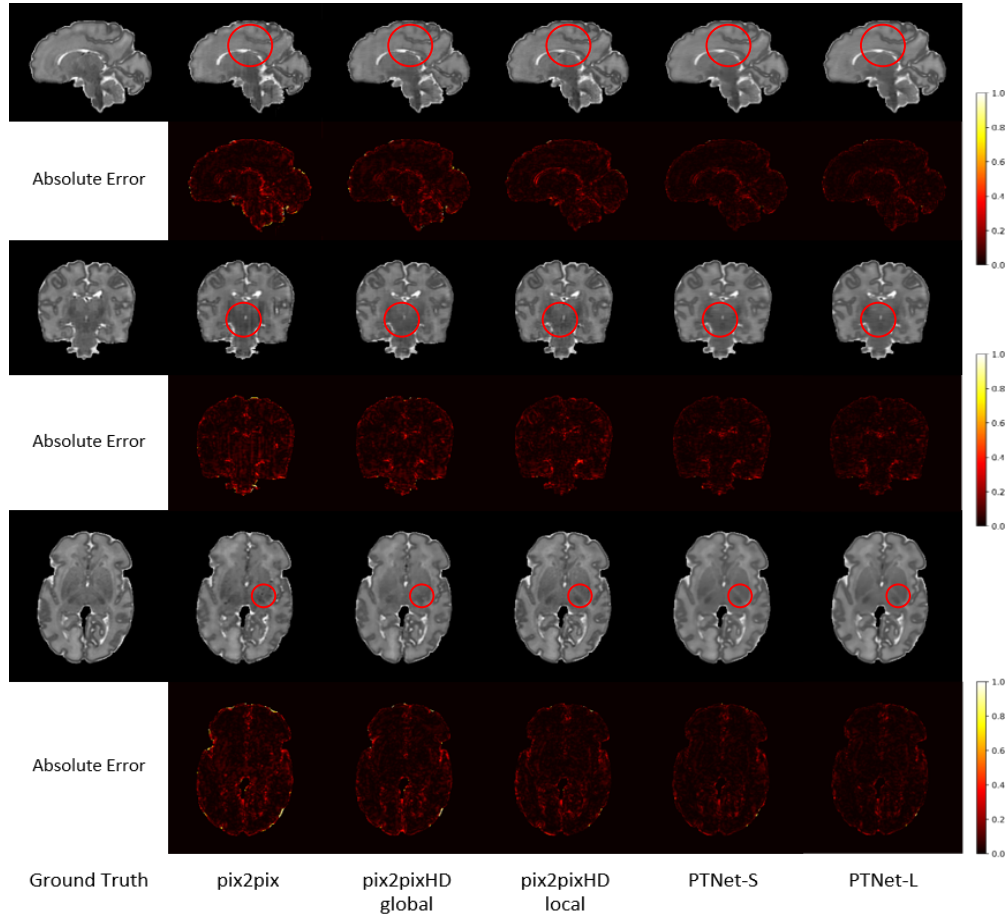

**Fig. 6.** MRI synthesis results from different models as well as their absolute error maps. From top to bottom: sagittal, coronal, and axial view. Absolute error is calculated based on normalized ground truth and converted scans, ranges from [0,1].

pix2pix model exhibited some patterns of artifacts.

### B. Quantitative Results on dHCP Dataset

Then we quantitatively compared results from our proposed PTNet with results from pix2pix series models. The mean values of SSIMs and pSNRs were calculated from the test dataset and are reported in **Table 2**. Both SSIMs and pSNRs of PTNet are significantly higher than those of pix2pix series models at $p < 0.05$. For T1w-to-T2w synthesis, PTNet-L provided a 2.83% and 4.55 dB improvement over the pix2pix for SSIM and pSNR, respectively. Similarly, for T2w-to-T1w synthesis, PTNet-L provided 2.13% and 2.33 dB improvement over pix2pix for SSIM and pSNR.

When compared with the largest model pix2pixHD-local, which had similar execution speed and also incorporated a pyramid design, PTNet-L on average improved the SSIM and pSNR by 1.19% and 1.50 dB for T1w-to-T2w and T2w-to-T1w. **Fig. 6**). No significant difference between PTNet-S and PTNet-L was observed.

**TABLE 2.** QUANTITATIVE RESULTS

| Models | SSIM (%) | pSNR (dB) | Params (M) | FPS (Hz) |
|---|---|---|---|---|
| T1w-to-T2w | | | | |
| pix2pix | 94.58 | 27.01 | 16.65 | 157 |
| pix2pixHD-global | 95.48 | 28.57 | 182.43 | 63 |
| pix2pixHD-local | 96.37 | 29.92 | 504.20 | 38 |
| **PTNet-S** | 97.30 | 31.48 | 8.78 | 37 |
| **PTNet-L** | **97.41** | **31.56** | 27.69 | 27 |
| T2w-to-T1w | | | | |
| pix2pix | 94.12 | 26.29 | 16.65 | 157 |
| pix2pixHD-global | 94.89 | 26.87 | 182.43 | 63 |
| pix2pixHD-local | 94.92 | 27.27 | 504.20 | 38 |
| **PTNet-S** | 96.17 | 28.56 | 8.78 | 37 |
| **PTNet-L** | **96.25** | **28.62** | 27.69 | 27 |

Bold indicated the best performance. Average results from 83 volumes were reported.

In addition to SSIM and pSNR, we considered two other factors including parameters number of each model (in million **M**) and frames per second (Hz). Plots of SSIM, pSNR, model size, and execution time are provided in **Fig.7**. The execution time (FPS in Hz) was denoted by the size of circles – the bigger the circle, the shorter of execution time. And an FPS around 30 was considered sufficient / good for MRI synthesis. Although pix2pix provided the fastest execution time (processes 157 slices per second), significant artifacts and loss of details were constantly observed in all three anatomical views (circled in Fig. 6)

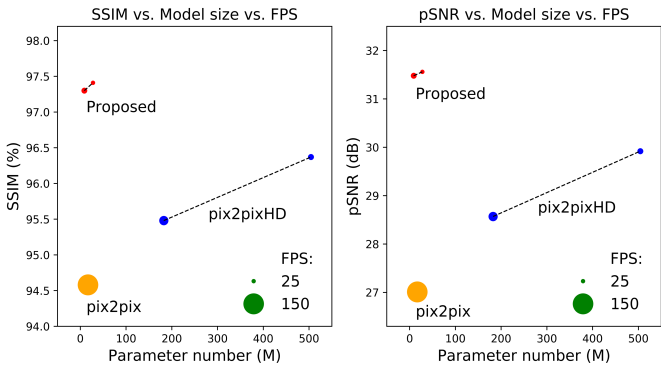

**Fig. 7**. Comparing PTNet with pix2pix and pix2pixHD models. Right: structural similarity index measure (SSIM) vs. model's parameter number and executing frame rate (FPS); Left: peak signal-to-noise ratio (pSNR) vs. parameter number and FPS. The execution time is calculated using a single NVIDIA GeForce RTX 2080 Ti GPU.

### C. Visual Comparisons of Models' Feature Maps

We further visualized and compared internal feature maps from pix2pixHD-global and PTNet-L. **Fig. 8** shows feature maps from the decoder path of PTNet-L with a matrix size of 56x64 (**panel a**) and feature maps from pix2pixHD-global with a matrix size of 56x64 (**panel b**). It was remarkable that the transformer-based network generated more structured activations given the same input. We speculate such a remarkable difference in feature maps may account for the improved quality of synthesized MRI scans.

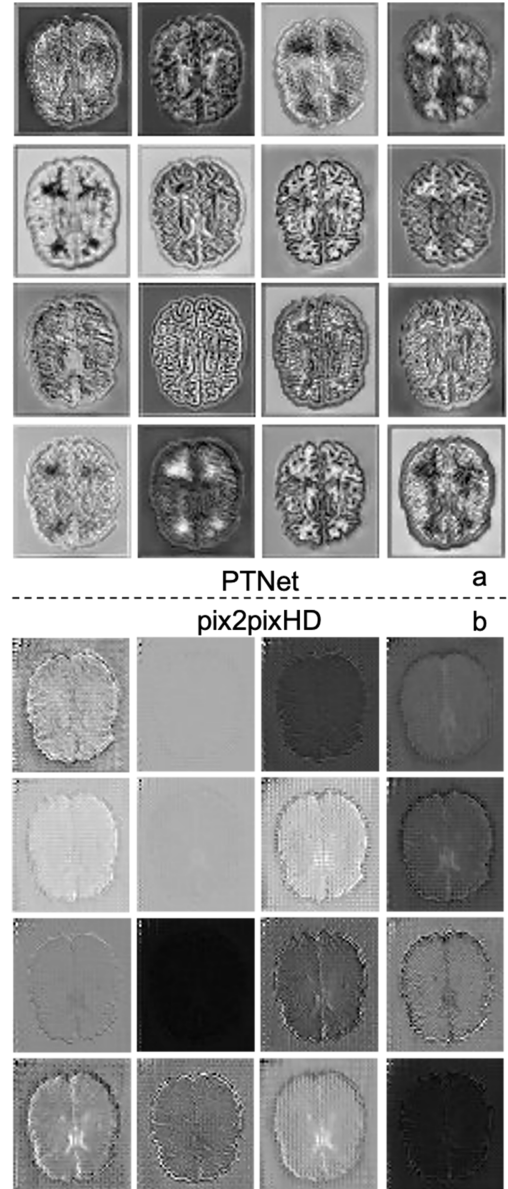

**Fig. 8**. Feature maps from the decoder path of pix2pixHD-global and proposed PTNet-L. a): Feature maps from the proposed PTN-L, with a size of 56x64. b): Feature maps from pix2pixHD-global model, with a size of 56x64.

### D. Applications - Synthesizing Corrupted Scans

To showcase potential utility in research, we directly applied our PTNet framework to synthesize T1w scans using good-quality T2w scans in our dataset. 43 T1w scans (9.37%) in our dataset have severe quality issues caused by factors such as motion and other scanner-related artifacts. In **Fig. 9,** we displayed three examples of synthesizing corrupted T1w scans using good quality T2w scans**.** Results demonstrated our proposed framework can be used to synthesize MRI scans to surrogate the corrupted ones, and ultimately increase the sample size.

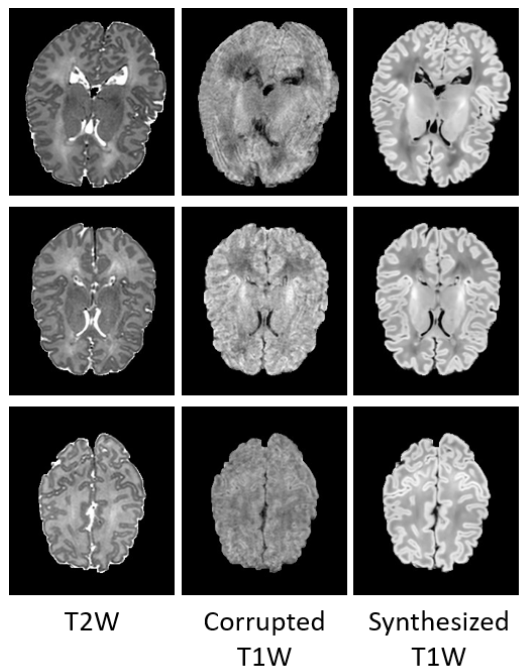

**Fig. 9**. Examples of synthesizing corrupted T1w scans using good quality T2w scans.

## V. Discussions

In this work, we introduced a novel MRI synthesis framework – PTNet, compared its performance with other commonly used models in a large infant MRI dataset, and explored applications of utilizing PTNet to synthesize realistic MRI scans to surrogate corrupted ones. PTNet consists of transformer layers, skip-connections, and multi-scale pyramid representation. To balance computing need and model accuracy, we incorporated the performer, a variant of the transformer architecture, with the original transformer to design a convolution-free, pyramid-like framework, which was experimentally proven to outperform the CNN-based pix2pix and pix2pixHD models in terms of conversion quality and model size. Overall, PTNet generated more realistic MRI scans with less motion- and scanner-related artifacts. In addition to synthesis improvement, PTNet still has a reasonable and practical execution time, which can process in around 30 slices per second.

Though we have evaluated our model in a large-scale MRI dataset, we cannot ignore several limitations in this work. First, only newborns' MRI scans were used here, thus we do not know if our model will have similar performance when tested in another independent datasets with infant MRI scans at different ages (e.g., 3m, 6m, 9m). Second, we did not observe significant difference between PTNet-S and PTNet-L in terms of performance, but their distinction might be found in other datasets, e.g., a longitudinal dataset. Therefore, further study needs to investigate their differences and extend their applications to other scenarios / tasks, e.g., image segmentation and registration.

## Acknowledgment

We thank the funding support from Environmental influences on Child Health Outcomes (ECHO) Program- Opportunities and Infrastructure Fund (EC0360, Yun Wang), along with National Institutes of Health (UH3 OD023328; R01 MH119510; R01 MH121070, Jonathan Posner).

## References

[1] S. J. Short *et al*., "Associations between white matter microstructure and infants' working memory," (in English), *Neuroimage,* vol. 64, pp. 156-166, Jan 1 2013, doi: 10.1016/j.neuroimage.2012.09.021.

[2] M. N. Spann, R. Bansal, T. S. Rosen, and B. S. Peterson, "Morphological Features of the Neonatal Brain Support Development of Subsequent Cognitive, Language, and Motor Abilities," (in English), *Human Brain Mapping,* vol. 35, no. 9, pp. 4459-4474, Sep 2014, doi: 10.1002/hbm.22487.

[3] A. M. Fjell *et al*., "Multimodal imaging of the self-regulating developing brain," (in English), *P Natl Acad Sci USA,* vol. 109, no. 48, pp. 19620-19625, Nov 27 2012, doi: 10.1073/pnas.1208243109.

[4] J. O'Muircheartaigh *et al*., "White Matter Development and Early Cognition in Babies and Toddlers," (in English), *Human Brain Mapping,* vol. 35, no. 9, pp. 4475-4487, Sep 2014, doi: 10.1002/hbm.22488.

[5] M. Prastawa, J. H. Gilmore, W. Lin, and G. Gerig, "Automatic segmentation of MR images of the developing newborn brain," *Med Image Anal,* vol. 9, no. 5, pp. 457-66, Oct 2005, doi: 10.1016/j.media.2005.05.007.

[6] J. H. Gilmore, R. C. Knickmeyer, and W. Gao, "Imaging structural and functional brain development in early childhood," *Nature Reviews Neuroscience,* vol. 19, no. 3, pp. 123-137, 2018, doi: 10.1038/nrn.2018.1.

[7] A. Makropoulos *et al*., "The developing human connectome project: A minimal processing pipeline for neonatal cortical surface reconstruction," *Neuroimage,* vol. 173, pp. 88-112, Jun 2018, doi: 10.1016/j.neuroimage.2018.01.054.

[8] L. Zollei, J. E. Iglesias, Y. Ou, P. E. Grant, and B. Fischl, "Infant FreeSurfer: An automated segmentation and surface extraction pipeline for T1-weighted neuroimaging data of infants 0-2 years," *Neuroimage,* vol. 218, p. 116946, Sep 2020, doi: 10.1016/j.neuroimage.2020.116946.

[9] S. U. Dar, M. Yurt, L. Karacan, A. Erdem, E. Erdem, and T. Cukur, "Image Synthesis in Multi-Contrast MRI With Conditional Generative Adversarial Networks," *IEEE Transactions on Medical Imaging,* vol. 38, no. 10, pp. 2375-2388, 2019, doi: 10.1109/tmi.2019.2901750.

[10] A. Jog, A. Carass, D. L. Pham, and J. L. Prince, "Random forest flair reconstruction from T1, T2, and PD-weighted MRI," (in eng), *Proc IEEE Int Symp Biomed Imaging,* vol. 2014, pp. 1079-1082, 2014, doi: 10.1109/ISBI.2014.6868061.

[11] J. E. Iglesias, E. Konukoglu, D. Zikic, B. Glocker, K. Van Leemput, and B. Fischl, "Is synthesizing MRI contrast useful for inter-modality analysis?," (in eng), *Medical image computing and computer-assisted intervention : MICCAI ... International Conference on Medical Image Computing and Computer-Assisted Intervention,* vol. 16, no. Pt 1, pp. 631-638, 2013, doi: 10.1007/978-3-642-40811-3_79.

[12] A. Jog, A. Carass, S. Roy, D. L. Pham, and J. L. Prince, "MR image synthesis by contrast learning on neighborhood ensembles," (in eng), *Medical image analysis,* vol. 24, no. 1, pp. 63-76, 2015, doi: 10.1016/j.media.2015.05.002.

[13] N. Burgos *et al*., "Attenuation Correction Synthesis for Hybrid PET-MR Scanners: Application to Brain Studies," *IEEE Transactions on Medical Imaging,* vol. 33, no. 12, pp. 2332-2341, 2014, doi: 10.1109/TMI.2014.2340135.

[14] M. I. Miller, G. E. Christensen, Y. Amit, and U. Grenander, "Mathematical textbook of deformable neuroanatomies," (in eng), *Proc Natl Acad Sci U S A,* vol. 90, no. 24, pp. 11944-11948, 1993, doi: 10.1073/pnas.90.24.11944.

[15] S. Roy, A. Carass, and J. L. Prince, "Magnetic Resonance Image Example-Based Contrast Synthesis," (in eng), *IEEE transactions on medical imaging,* vol. 32, no. 12, pp. 2348-2363, 2013, doi: 10.1109/TMI.2013.2282126.

[16] S. Roy, A. Carass, and J. Prince, "A Compressed Sensing Approach for MR Tissue Contrast Synthesis," in *Information Processing in Medical Imaging*, Berlin, Heidelberg, G. Székely and H. K. Hahn, Eds., 2011// 2011: Springer Berlin Heidelberg, pp. 371-383.

[17] A. Jog, A. Carass, S. Roy, D. L. Pham, and J. L. Prince, "Random forest regression for magnetic resonance image synthesis," *Medical Image Analysis,* vol. 35, pp. 475-488, 2017, doi: 10.1016/j.media.2016.08.009.

[18] A. Jog, S. Roy, A. Carass, and J. L. Prince, "Magnetic resonance image synthesis through patch regression," in *2013 IEEE 10th International Symposium on Biomedical Imaging*, 7-11 April 2013 2013, pp. 350-353, doi: 10.1109/ISBI.2013.6556484.

[19] A. Chartsias, T. Joyce, M. V. Giuffrida, and S. A. Tsaftaris, "Multimodal MR Synthesis via Modality-Invariant Latent Representation," *IEEE Transactions on Medical Imaging,* vol. 37, no. 3, pp. 803-814, 2018, doi: 10.1109/TMI.2017.2764326.

[20] I. J. Goodfellow *et al*., "Generative adversarial nets," presented at the Proceedings of the 27th International Conference on Neural Information Processing Systems - Volume 2, Montreal, Canada, 2014.

[21] T. Karras, T. Aila, S. Laine, and J. Lehtinen, "Progressive Growing of GANs for Improved Quality, Stability, and Variation," *ICLR,* 2018 2018.

[22] T.-C. Wang, M.-Y. Liu, J.-Y. Zhu, A. Tao, J. Kautz, and B. Catanzaro, "High-Resolution Image Synthesis and Semantic Manipulation with Conditional GANs," in *2018 IEEE/CVF Conference on Computer Vision and Pattern Recognition*, June 2018 2018, pp. 8798-8807, doi: 10.1109/CVPR.2018.00917.

[23] P. Isola, J. Zhu, T. Zhou, and A. A. Efros, "Image-to-Image Translation with Conditional Adversarial Networks," in *2017 IEEE Conference on Computer Vision and Pattern Recognition (CVPR)*, 21-26 July 2017 2017, pp. 5967-5976, doi: 10.1109/CVPR.2017.632.

[24] Q. Yang, N. Li, Z. Zhao, X. Fan, E. I. C. Chang, and Y. Xu, "MRI Cross-Modality Image-to-Image Translation," (in eng), *Scientific reports,* vol. 10, no. 1, pp. 3753-3753, 2020, doi: 10.1038/s41598-020-60520-6.

[25] A. Vaswani *et al*., "Attention is all you need," presented at the Proceedings of the 31st International Conference on Neural Information Processing Systems, Long Beach, California, USA, 2017.

[26] A. Dosovitskiy *et al*., "An Image is Worth 16x16 Words: Transformers for Image Recognition at Scale," *arXiv pre-print server,* 2020-10-22 2020, doi: None arxiv:2010.11929.

[27] K. Choromanski *et al*., "Rethinking Attention with Performers," *arXiv pre-print server,* 2021-03-09 2021, doi: None arxiv:2009.14794.

[28] L. Yuan *et al*., "Tokens-to-Token ViT: Training Vision Transformers from Scratch on ImageNet," *arXiv pre-print server,* 2021-03-22 2021, doi: None arxiv:2101.11986.

[29] O. Ronneberger, P. Fischer, and T. Brox, "U-Net: Convolutional Networks for Biomedical Image Segmentation," in *Medical Image Computing and Computer-Assisted Intervention – MICCAI 2015*, Cham, N. Navab, J. Hornegger, W. M. Wells, and A. F. Frangi, Eds., 2015// 2015: Springer International Publishing, pp. 234-241.

[30] A. Makropoulos *et al*., "The developing human connectome project: A minimal processing pipeline for neonatal cortical surface reconstruction," *NeuroImage,* vol. 173, pp. 88-112, 2018, doi: 10.1016/j.neuroimage.2018.01.054.

[31] Z. Liu *et al*., "Swin Transformer: Hierarchical Vision Transformer using Shifted Windows," *arXiv pre-print server,* 2021-03-25 2021, doi: None arxiv:2103.14030.

[32] D. A. Hudson and C. L. Zitnick, "Generative Adversarial Transformers," *arXiv pre-print server,* 2021-03-02 2021, doi: None arxiv:2103.01209.

[33] A. Vaswani *et al*., "Attention is all you need," *arXiv preprint arXiv:1706.03762,* 2017.

[34] X. Huang, Y. Li, O. Poursaeed, J. Hopcroft, and S. Belongie, "Stacked Generative Adversarial Networks," in *2017 IEEE Conference on Computer Vision and Pattern Recognition (CVPR)*, 21-26 July 2017 2017, pp. 1866-1875, doi: 10.1109/CVPR.2017.202.

[35] P. Burt and E. Adelson, "The Laplacian Pyramid as a Compact Image Code," *IEEE Transactions on Communications,* vol. 31, no. 4, pp. 532-540, 1983, doi: 10.1109/TCOM.1983.1095851.

[36] Brown and Lowe, "Recognising panoramas," in *Proceedings Ninth IEEE International Conference on Computer Vision*, 13-16 Oct. 2003 2003, pp. 1218-1225 vol.2, doi: 10.1109/ICCV.2003.1238630.

[37] E. Denton, S. Chintala, A. Szlam, and R. Fergus, "Deep generative image models using a Laplacian pyramid of adversarial networks," presented at the Proceedings of the 28th International Conference on Neural Information Processing Systems - Volume 1, Montreal, Canada, 2015.

[38] J. Johnson, A. Alahi, and L. Fei-Fei, "Perceptual Losses for Real-Time Style Transfer and Super-Resolution," in *Computer Vision – ECCV 2016*, Cham, B. Leibe, J. Matas, N. Sebe, and M. Welling, Eds., 2016// 2016: Springer International Publishing, pp. 694-711.

[39] A. Horé and D. Ziou, "Image Quality Metrics: PSNR vs. SSIM," in *2010 20th International Conference on Pattern Recognition*, 23-26 Aug. 2010 2010, pp. 2366-2369, doi: 10.1109/ICPR.2010.579.

[40] W. Zhou, A. C. Bovik, H. R. Sheikh, and E. P. Simoncelli, "Image quality assessment: from error visibility to structural similarity," *IEEE Transactions on Image Processing,* vol. 13, no. 4, pp. 600-612, 2004, doi: 10.1109/TIP.2003.819861.